\documentclass[
 preprint,
  prb,
  showpacs,
  amsmath,
  amssymb,
  superscriptaddress,
  floatfix,
  endfloats
]{revtex4}

\usepackage{bm}
\usepackage{dsfont}
\usepackage{graphicx}
\usepackage{pifont}
\usepackage{textcomp}
\usepackage{gensymb}
\usepackage{accents}
\usepackage{upgreek}
\usepackage{natbib}
\usepackage{tabularx}
 
\renewcommand{\cite}[1]{{[}\onlinecite{#1}{]}}
\newcommand{\be}{\begin{equation}}
\newcommand{\e}{\end{equation}}

\newcommand{\pa}{\partial} 
 
\newcommand{\beml}{\begin{subequations}} 
\newcommand{\eml}{\end{subequations}} 
\newcommand{\beq}{\begin{eqnarray}} 
\newcommand{\eq}{\end{eqnarray}} 
\newcommand{\ba}{\begin{array}} 
\newcommand{\ea}{\end{array}} 
\newcommand{\bpm}{\begin{pmatrix}} 
\newcommand{\epm}{\end{pmatrix}} 
\newcommand{\bc}{\begin{cases}} 
\newcommand{\ec}{\end{cases}} 
\newcommand{\lt}{\left} 
\newcommand{\rt}{\right} 
\newcommand{\n}{\nonumber}

\newcommand{\ep}{\varepsilon} 
 
\newcommand{\bb}{\boldsymbol}

\begin{document}
	
\title{Magnetoresistance in the in-plane magnetic field induced semi-metallic phase of inverted HgTe quantum wells}

\author{T.~Khouri}
\email{Thomas.Khouri@ru.nl}
\affiliation{High Field Magnet Laboratory (HFML-EMFL), Radboud University, Toernooiveld 7, 6525 ED Nijmegen, The Netherlands} 
\affiliation{Radboud University, Institute of Molecules and Materials, Heyendaalseweg 135, 6525 AJ Nijmegen, Netherlands} 
 
\author{S.~Pezzini}
\affiliation{High Field Magnet Laboratory (HFML-EMFL), Radboud University, Toernooiveld 7, 6525 ED Nijmegen, The Netherlands} 
\affiliation{Radboud University, Institute of Molecules and Materials, Heyendaalseweg 135, 6525 AJ Nijmegen, Netherlands} 
 
\author{M.~Bendias}
\affiliation{Physikalisches Institut (EP3), Universit\"at W\"urzburg, Am Hubland, 97074 W\"urzburg,Germany}
\homepage[Orcid ID M.~Bendias: ]{https://orcid.org/0000-0002-9551-0454}

\author{P.~Leubner}
\affiliation{Physikalisches Institut (EP3), Universit\"at W\"urzburg, Am Hubland, 97074 W\"urzburg,Germany}
\affiliation{Department of Applied Physics, Eindhoven University of Technology (TU/e) , 5600 MB Eindhoven, the Netherlands}

\author{U.~Zeitler}
\affiliation{High Field Magnet Laboratory (HFML-EMFL), Radboud University, Toernooiveld 7, 6525 ED Nijmegen, The Netherlands} 
\affiliation{Radboud University, Institute of Molecules and Materials, Heyendaalseweg 135, 6525 AJ Nijmegen, Netherlands} 

\author{N.~E.~Hussey}
\affiliation{High Field Magnet Laboratory (HFML-EMFL), Radboud University, Toernooiveld 7, 6525 ED Nijmegen, The Netherlands} 
\affiliation{Radboud University, Institute of Molecules and Materials, Heyendaalseweg 135, 6525 AJ Nijmegen, Netherlands} 

\author{H.~Buhmann}
\affiliation{Physikalisches Institut (EP3), Universit\"at W\"urzburg, Am Hubland, 97074 W\"urzburg,Germany}

\author{L.~W.~Molenkamp}
\affiliation{Physikalisches Institut (EP3), Universit\"at W\"urzburg, Am Hubland, 97074 W\"urzburg,Germany}

\author{M.~Titov}
\affiliation{Radboud University, Institute of Molecules and Materials, Heyendaalseweg 135, 6525 AJ Nijmegen, Netherlands} 
\affiliation{ITMO University, Saint Petersburg 197101, Russia}

\author{S.~Wiedmann}
\email{Steffen.Wiedmann@ru.nl}
\affiliation{High Field Magnet Laboratory (HFML-EMFL), Radboud University, Toernooiveld 7, 6525 ED Nijmegen, The Netherlands} 
\affiliation{Radboud University, Institute of Molecules and Materials, Heyendaalseweg 135, 6525 AJ Nijmegen, Netherlands}

\date{\today}
	
\begin{abstract}

In this study we have measured the magnetoresistance response of inverted HgTe quantum wells in the presence of a large parallel magnetic field up to 33 T is applied. We show that in quantum wells with inverted band structure a monotonically decreasing magnetoresistance is observed when a magnetic field up to order 10 T is applied parallel to the quantum well plane.
This feature is accompanied by a vanishing of non-locality and is consistent with a predicted modification of the energy spectrum that becomes gapless at a critical in-plane field $B_{c}$.
Magnetic fields in excess of $B_c$ allow us to investigate the evolution of the magnetoresistance in this field-induced semi-metallic region beyond the known regime. After an initial saturation phase in the presumably gapless phase, we observe a strong upturn of the longitudinal resistance. A small residual Hall signal picked up in non-local measurements suggests that this feature is likely a bulk phenomenon and caused by the semi-metallicity of the sample. Theoretical calculations indeed support that the origin of these features is classical and a power law upturn of the resistance can be expected due to the specifics of two-carrier transport in thin (semi-)metallic samples subjected to large magnetic fields.
\end{abstract}
	
\maketitle

Magnetoresistance studies represent one of the most powerful tools to investigate the electronic structure of conducting materials.
In a perpendicular magnetic field, the orbital motion of charge carriers creates a variety of classical and quantum Hall effects, along with Shubnikov-de Haas oscillations that are routinely observed in  two-dimensional electron (hole) gases. These effects are useful in determining the electron (or hole) concentration, the mobility of charge carriers as well as the spectrum of Landau levels. 
While this perpendicular magnetic field configuration is widely employed in a variety of measurements, in-plane magnetic fields are much less commonly explored as orbital effects are, generally, assumed to be absent and electrons move under the competing influence of the Lorentz force and confinement potential. Any finite magnetoresistance in parallel fields is therefore often claimed to be a signature of spin effects. While this would, indeed, be true in an ideal two-dimensional (2D) system with zero thickness and vanishing spin-orbit coupling \cite{PhysRevB.94.085302}, the situation may be more complex in quasi-2D structures especially those where both electron and hole carriers coexist \cite{Alekseev2015,Alekseev2017}. 

Quasi-2D systems usually have a confinement potential with an effective width of a few tens of nanometers. This length scale becomes rapidly comparable to the magnetic length $l_B=\sqrt{\hbar/eB}$  for a magnetic field of just a few Tesla. In this case, strong magneto-orbital effects can lead to magnetoresistance features \cite{Zhou2010} and even cause an effective 2D to 3D crossover provided the energy of the orbital motion exceeds the sub-band separation \cite{DasSarma2000}. In addition to the orbital effects, the in-plane magnetic field may strongly affect the screening of charged impurities, thus leading to an effective increase of disorder scattering that can enhance localization effects and induce a phase transition to an insulating state \cite{DasSarma2014,Piot2009}.
 While most of these phenomena have been studied in conventional electron systems and are only observable at temperatures at or below $\sim$ 1 K, it has been recently demonstrated that in the topological regime of an 8 nm HgTe quantum well grown in the [013] direction, there is a strong monotonic decrease of the longitudinal resistance accompanied by a vanishing non-local signal  \cite{Gusev2011b,Gusev2013f} - the  hallmark of a 2D TI \cite{roth_nonlocal_2009}. These features are consistent with a theoretically predicted magnetic-field driven phase transition from a 2D TI state to a gapless (semi-)metallic state \cite{Raichev2012a}.

In this work, we demonstrate that similar features are observable in HgTe quantum wells grown in the more symmetric [001] direction (as compared to the so far studied [013] direction) with quantum well thicknesses between 7 and 11 nm that possess an inverted band structure. We further investigate the evolution of the magnetoresistance up to 33 T by means of magneto-transport utilizing standard lock-in technology.
At low magnetic fields, we observe a negative magnetoresistance that likely indicates the predicted phase transition from a 2D TI to a semi-metallic phase. With a further increase of magnetic field the magnetoresistance saturates in the supposedly gapless semi-metallic phase before appearing to diverge at the highest fields of 33 T.
Although such a strong magnetoresistance may be indicative of a new high field phase transition, we are able to exclude this possibility from the presence of a small Hall component picked up in non-local measurements that suggests a gapless bulk state in agreement with the theoretical calculations of Ref.\cite{Raichev2012a}.
In semi-metallic systems the electron-hole recombination length sets up yet another length scale \cite{Alekseev2017} that can easily exceed the system width. In this case, a non-trivial magnetoresistance emerges naturally due to skin effects on the scale of the recombination length \cite{Rashba1976}.

\begin{figure}
\includegraphics[width=0.5\linewidth]{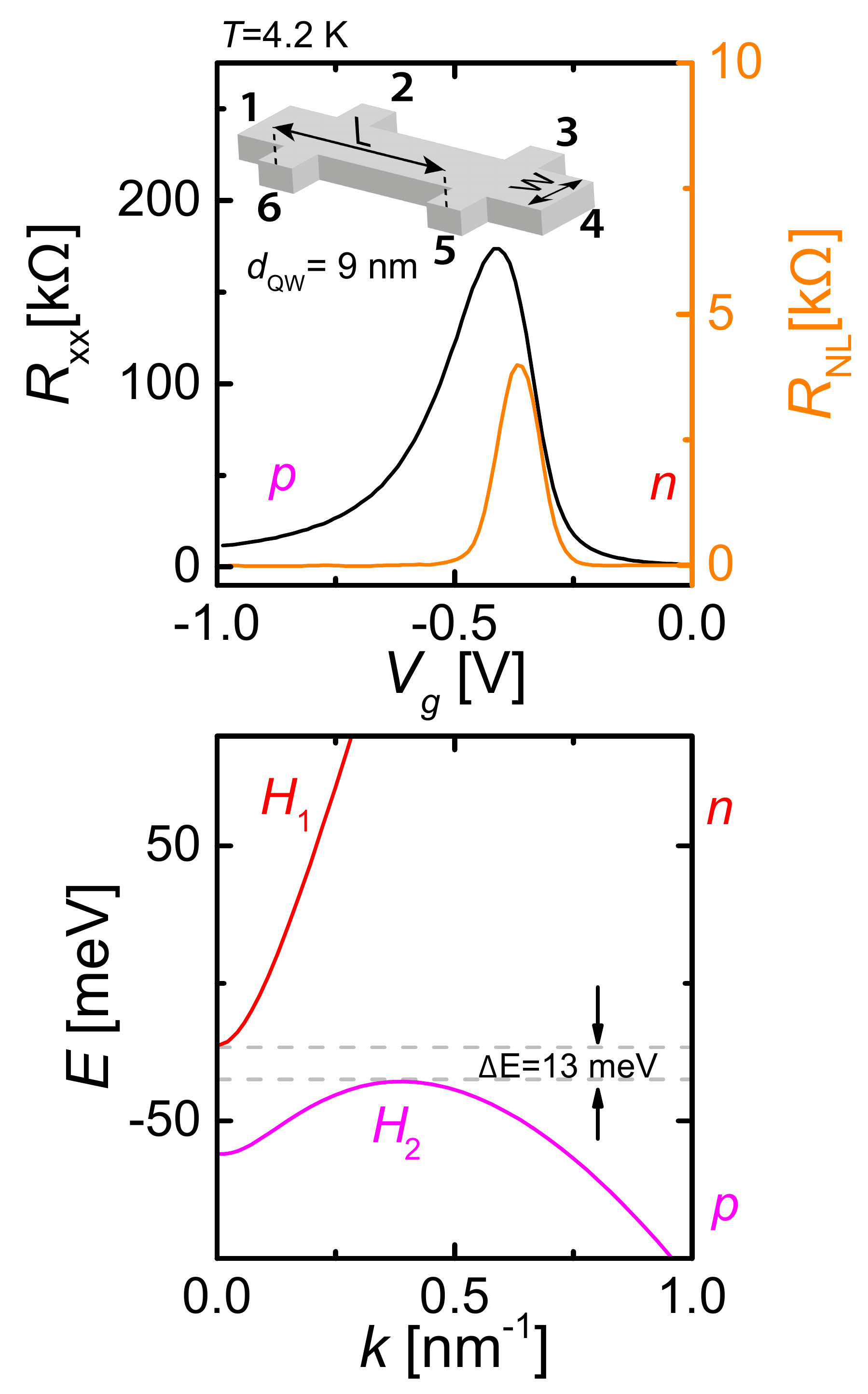}%
\caption{ 
a) Longitudinal (black, left axis) and non-local (orange, right axis) resistance measurements in a 9\,nm thick HgTe quantum well at zero magnetic field. The $n$ and $p$ conducting regions have been marked accordingly. The inset of the figure shows a schematic of the sample geometry. b) The results of the self-consistent band structure calculations based on the $8 \times 8$ $k\cdot p$ Kane model for a HgTe quantum well where the strain effects originating from the substrate are taken into account. Conduction and valence bands are labelled with $H_1$ and $E_1$, respectively. The indirect band gap of $\sim$ 13 meV is indicated by the grey dashed lines.
 }
 \label{Fig1} 
 \end{figure}   

Our samples are undoped HgTe quantum wells grown by molecular beam epitaxy (MBE) in the [001] direction that are structured into standard 6- or 8-terminal Hall bars with dimensions of $L\times W=600\times 200$~$\mu$m$^2$ and $1200\times 200$~$\mu$m$^2$ respectively. A sketch of the sample geometry is provided in the inset of Fig.\ref{Fig1}~a) where we label the electrical contacts from 1 to 6. In this study, we investigate samples with different quantum well thicknesses between 7 and 11\,nm that possess an inverted band structure but varying size of the bulk band gap. A summary of all samples can be found in table~\ref{Table1} of the appendix. In the following we will mainly focus on sample \# 3 with a quantum well thickness of 9 nm that is well inside the inverted (topologically non-trivial) regime. It should be noted, however, that all our samples show qualitatively the same behaviour in the magnetoresistance.  
Data from samples with quantum well thicknesses of 7, 7.5 and 11 nm are presented in the appendix. All samples are equipped with a top-gate that allows for a precise control of the charge carrier concentration and thus allow us to study an extended area around the band gap region. 
For each sample, we perform self-consistent $\mathbf{k\cdot p}$ calculations of the energy spectrum \cite{Novik2005} where strain effects of the substrate \cite{Leubner2016} are taken into account. The corresponding band structure is presented in Fig.\ref{Fig1}~b). Due to a valence band maximum at high $k$-values, the valence and conduction bands are separated by an indirect band gap of $\sim$ 13~meV in which a 2D TI state should become observable. 
 \begin{figure}
 \includegraphics[width=\linewidth]{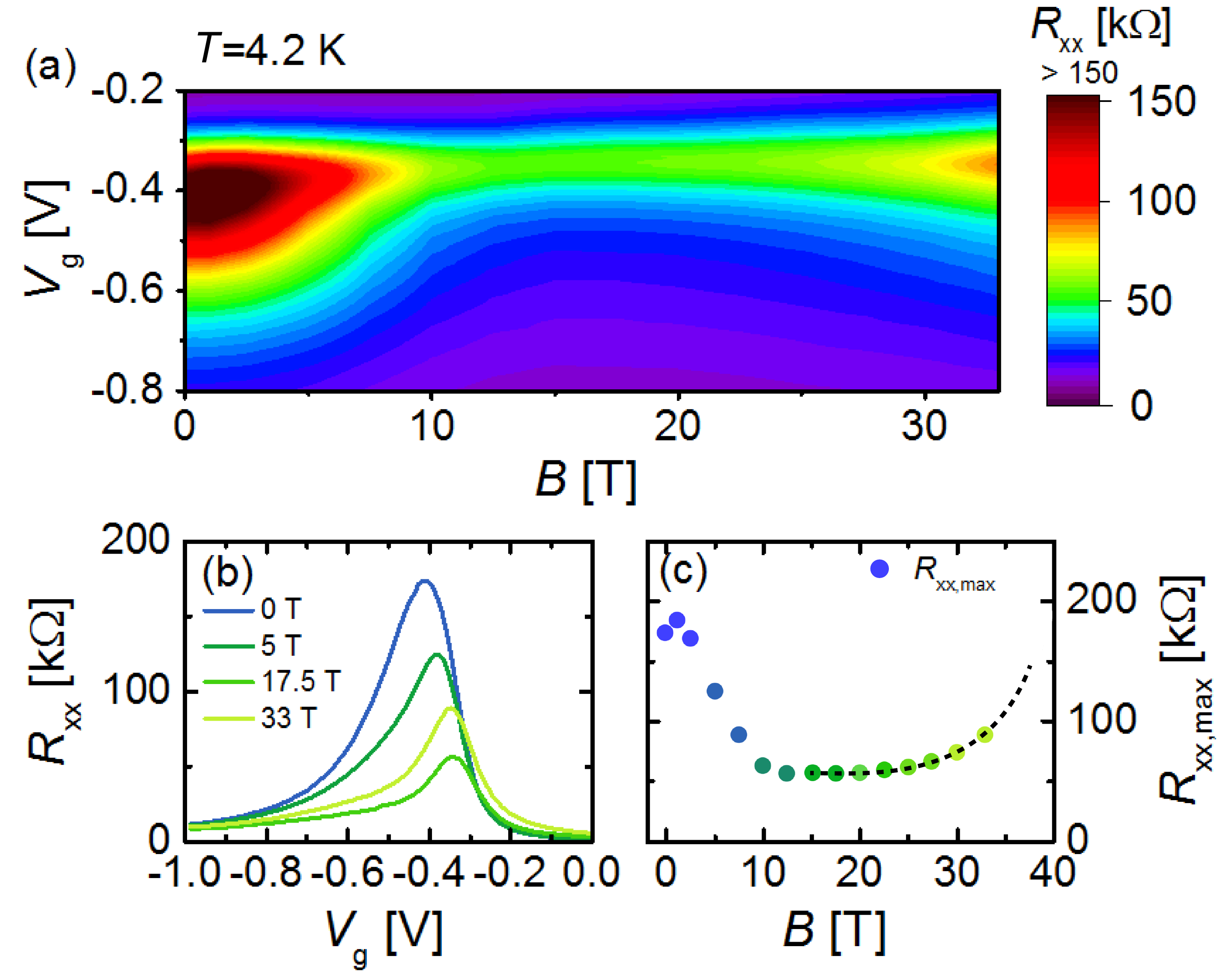}%
 \caption{
a) 2D false colour plot of the longitudinal resistance as a function of gate-voltage $V_{g}$ and in-plane magnetic field B at $T$=4.2 K. b) Gate-dependent measurements of $R_{xx}$ at fixed magnetic fields. Different magnetic fields are shown in different colours. The curves are selected to show the overall behaviour of the sample. c) Extracted peak resistance $R_{xx,max}$ as a function of the in-plane magnetic field up to 33 T. The black dashed line is a guide for the eye that shows the diverging resistance above 10 T. }
 \label{Fig2}%
 \end{figure}   
To verify this experimentally, we investigate the zero-field spectrum of our sample using a top-gate to tune the Fermi energy from the $n$-doped regime through the bulk band gap into the $p$-doped regime as shown in Fig.~\ref{Fig1}~a). In this figure,  we present low temperature measurements ($T$= 4.2 K) of the longitudinal $R_\textrm{xx}$ and the non-local resistance $R_{NL}$ of the 9 nm sample  at zero magnetic field. The longitudinal resistance is measured by passing electric current through the contacts 1 and 4, while the voltage drop is measured between the contacts 2,3 or 5,6. (Note that we only show one of the two possibilities as both configurations show the same qualitative behaviour with and without magnetic field.) 

The maximum of the longitudinal resistance $R_\textrm{xx,max}$ usually coincides with the Fermi energy being located in the middle of the bulk band gap. Due to the topological properties of the sample band structure, the current in this regime is expected to be carried by the helical edge modes. This can be verified by the presence of a large non-local resistance \cite{roth_nonlocal_2009} that vanishes outside of the band gap. We measure the non-local signal by spatially separating current (2 and 6) and voltage (3 and 5) contacts in a straightforward geometrical configuration to avoid possible stray effects of charge carriers.
From the measurements it is evident that both the value of the local and non-local resistances do not match the expected values of an ideal 2D TI of 12.9 k$\Omega$ and 17.2 k$\Omega$, respectively. The discrepancy may be caused by disorder-state induced charge puddles that allow for backscattering events to occur \cite{vayrynen_helical_2013}, especially in samples with relatively large dimensions as investigated here.

 \begin{figure}
 \includegraphics[width=0.6\linewidth]{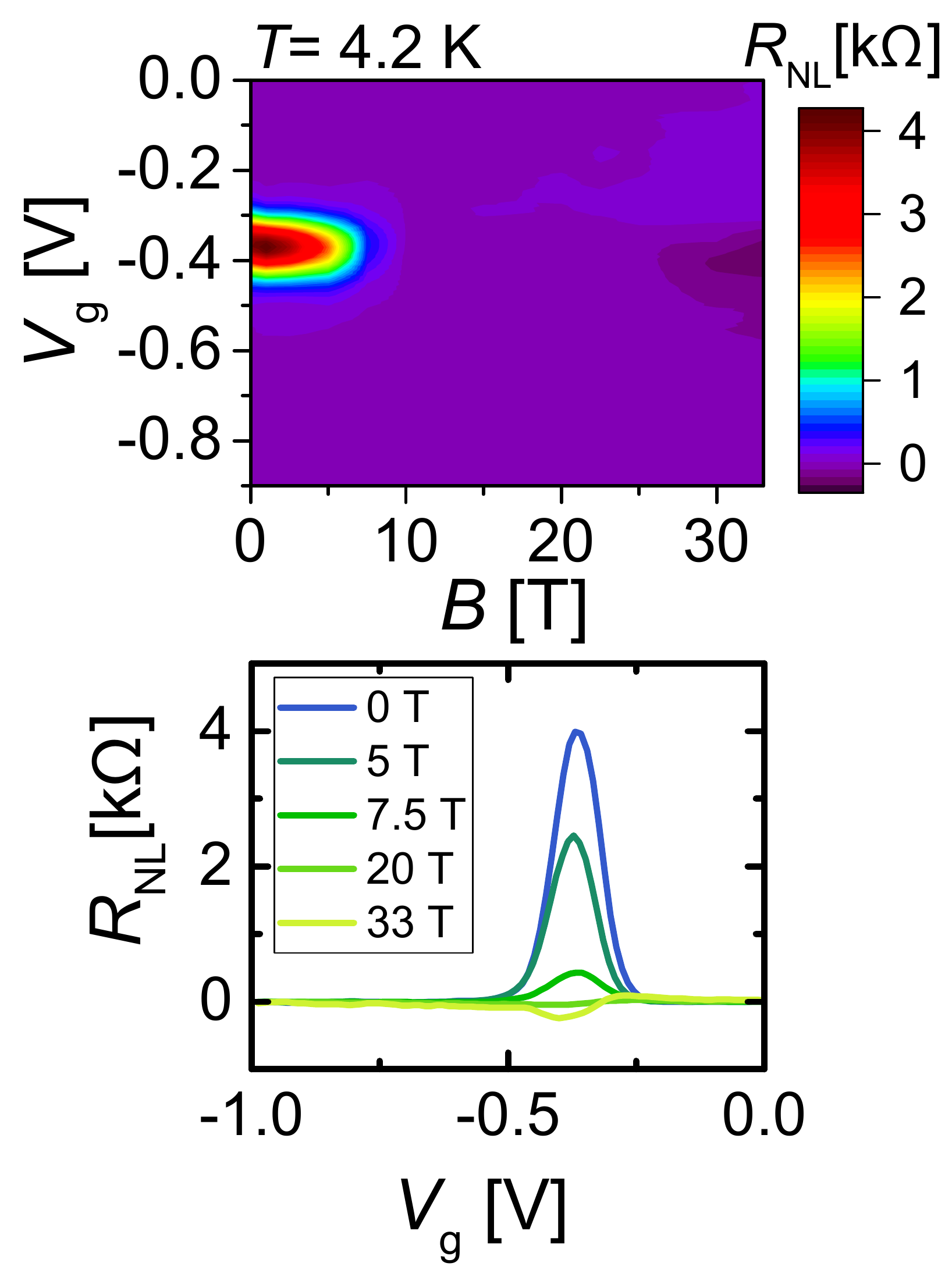}%
 \caption
 {a) 2D false colour plot of the non-local resistance as a function of gate-voltage $V_{g}$ and in-plane magnetic field $B$ at $T$=4.2 K. b)  Gate-dependent measurements of $R_{NL}$ at fixed magnetic fields. Different magnetic fields are shown in different colours. The selected curves show the overall behaviour of the sample.}
 \label{Fig3} 
 \end{figure}   
 
After experimentally verifying the presence of a 2D TI state at zero magnetic field, we apply an in-plane magnetic field by carefully aligning the quantum well plane to the magnetic field using an in-situ rotation stage. The alignment is achieved by  minimizing the Hall voltage of the sample at a well-defined electron concentration at $V_g=1$\,V. From a residual Hall signal (not shown here), we estimate the perpendicular component of the magnetic field to stay below 50 mT at a total magnetic field of 33 T.
In addition to the sample signal, we also monitored an external Hall probe glued to the back of the rotation stage to exclude any unwanted changes of angle during our measurements.

In the experiment, we measure the longitudinal resistance as a function of the gate voltage $V_g$ while increasing the value of the magnetic field in a stepwise fashion. This allows us to map an extended region around the band gap which should be affected by the in-plane magnetic field. The acquired data is summarized in a 2D colour plot presented in panel a) of Fig.~\ref{Fig2}. For a better comprehensive view, we additionally show in the panels b) and c) of Fig.~\ref{Fig2} different line-cuts reflecting the behaviour of the longitudinal resistance as a function of the gate voltage and magnetic field, respectively. 

 \begin{figure}
 \includegraphics[width=\linewidth]{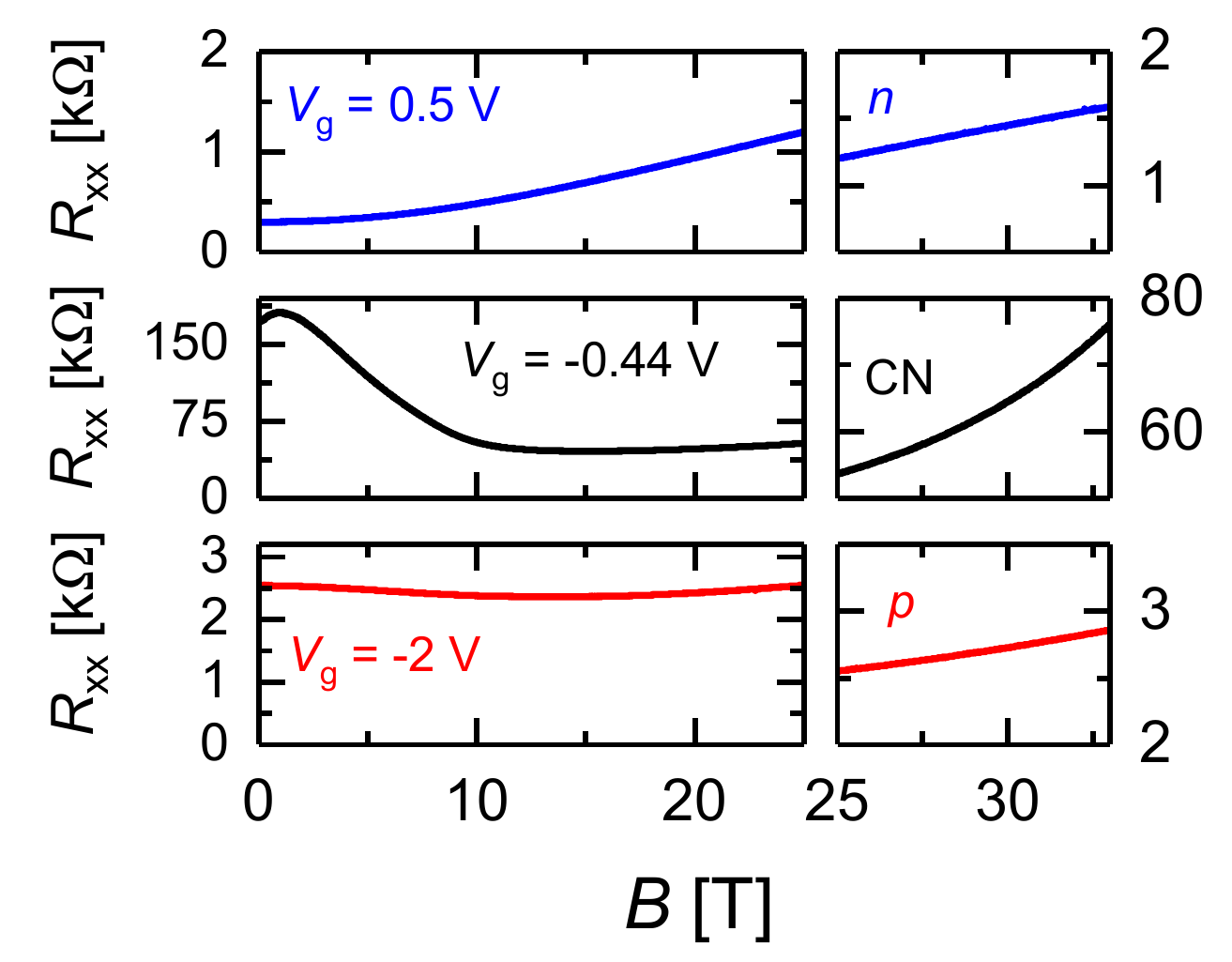}%
 \caption{Magnetoresistance of the 9\,nm sample at finite electron ($n$, top) and hole ($p$, bottom) concentrations as well as close to charge neutrality (CN, middle). The figure is divided into two parts where on the left, the magnetoresistance up to 25 T is shown and on the right a magnified view of the response in the highest fields up to 33 T is presented.}
 \label{Fig4} 
 \end{figure}

In the first part of our discussion, we focus on the region of the bulk band gap which is indicated by the peak resistance $R_\textrm{xx,max}$. For this purpose, we plot the measured values of $R_\textrm{xx,max}$ versus the in-plane magnetic field in Fig.~\ref{Fig2}~(c) (9\,nm sample) and in  Fig.~\ref{Fig5} of the appendix (7, 7.5 and 11 nm sample)

At low magnetic fields, we observe the same quantitative behaviour as reported in Ref.~\cite{Gusev2013b}. After an initial increase of $R_\textrm{xx,max}$ at $B\sim 1$\,T, a strong monotonic decrease of $R_\textrm{xx}$ occurs that eventually saturates at field strengths of the order of $10$\,T.  The feature at $B\approx 1$\,T is consistent with an increased backscattering from the 2D TI state caused by the breaking of time-reversal symmetry due to the magnetic field. A further increase is thus expected to lead to a positive magnetoresistance in contrast to our observations of a negative MR. This drop in the longitudinal resistance is accompanied by a vanishing of non-local signals, shown in Fig.~\ref{Fig3}. This indicates that charge transport no longer occurs through edge modes but is mainly carried by the bulk. Our observations are consistent with the previous interpretation of a phase transition from a 2D TI state to a gapless semi-metallic state.
We stress that while such a negative magnetoresistance may also be caused due to impurity scattering \cite{PhysRevB.94.081302,PhysRev.104.900}, we find  the magnetoresistance in HgTe to be isotropic, i.e. there is no difference in the response  for magnetic fields applied parallel or perpendicular to the current, making this possibility unlikely.
We would also like to point out that in the transient regime, the position of the longitudinal resistance peak is shifted and its width in gate-voltage terms shrinks considerably. Such effects are often attributed to undesired hysteresis effects of the top-gate \cite{Hinz2006} ,but in this experiment, they may also reflect a field-induced change in the sub-band structure that is reconstructed by the magnetic field. The exact relation between the gate voltage and the Fermi energy in our sample, however, remains too complicated to confirm this speculation.\\

When the magnetic field is increased further, we observe that the value of $R_\textrm{xx,max}$ of the 9 nm sample remains approximately constant  up to fields of order $B\sim 25$\,T. Above this field scale, however,  we see an unexpected upturn of the magnetoresistance, that has so far not been observed. This strong positive magnetoresistance is present in all our samples (independent of the quantum well width although its onset and strength varies (see Fig.~\ref{Fig2}~c) and panels a), c), and e) of Fig.~\ref{Fig5}).
This magnetoresistance is in fact strongest close to charge neutrality and weakens when the Fermi energy shifted far into the  $n$ ($V_g$= 0.5 V) or $p$-conducting ($V_g$= -2 V) regime as demonstrated in Fig.~\ref{Fig4}.
A large magnetoresistance is frequently observed in semi-metallic samples including topologically non-trivial materials such as Weyl semi-metals although its origin is still under debate. Apart from exotic topological explanations \cite{Niu2016,Tafti2016,Liang2015}, such a magnetoresistance may arise from a field-induced metal-insulator transition \cite{Tafti2016}, e.g. an opening of a band gap. This, however, contradicts the existing theoretical model for HgTe quantum wells in an in-plane magnetic field configuration where the bandstructure is expected to be gapless. This theoretical prediction is supported by the experimental observation of a clear change in parity of the non-local signal at 33~T that occurs close to charge neutrality. It is likely that this small feature is caused by the residual perpendicular component of the magnetic field that gives rise to a small Hall component which rises steeply in this low density regime. The presence of this well-defined Hall effect, however, indicates the presence of a bulk state exists, which is then presumably responsible for the observed magnetoresistance.
Indeed, a large unsaturating magnetoresistance has been shown to arise in compensated semi-metals close to perfect electron-hole compensation \cite{Luo2015,Pletikosic2014,Ali2014,FallahTafti2016}.
In a confined geometry or in the presence of macroscopic charge inhomogeneities, a power-law upturn of the magnetoresistance has been predicted to arise in a semi-metal phase close to charge neutrality \cite{Alekseev2015,Alekseev2017}.
To show that the observed magnetoresistance upturn in strong magnetic fields may be attributed to similar classical mechanisms in the specific case of a two-carrier transport in a thin film geometry, we adapt these existing models to our experimental conditions.

\begin{figure}
\includegraphics[width=\linewidth]{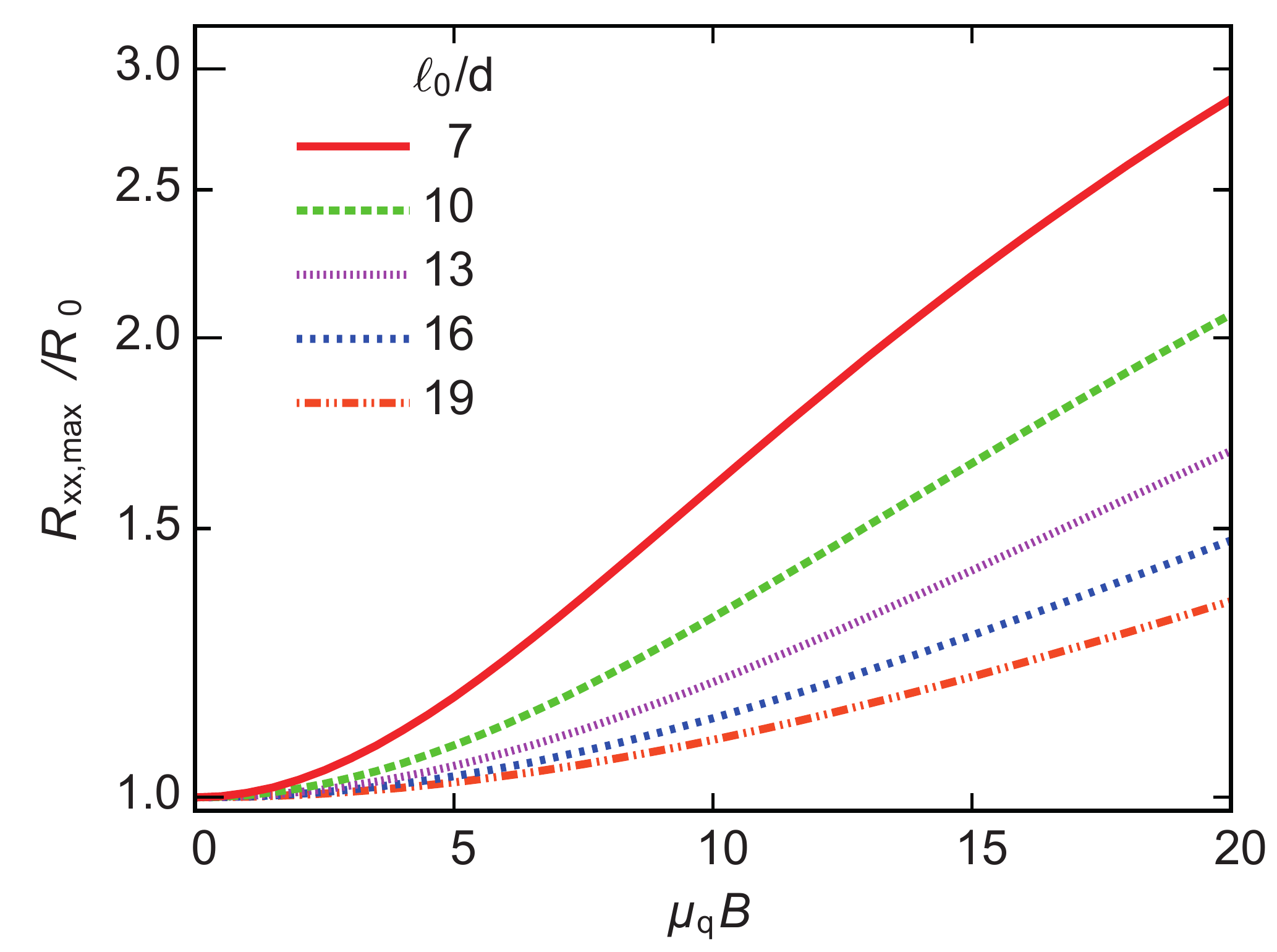}%
\caption{Classical magnetoresistance from Eq.~(\ref{theory}) versus  in-plane magnetic field $B$ directed perpendicular to the current in a thin three-dimensional sample with a sample thickness $d$ that is smaller than the zero-field electron-hole recombination length $\ell_0$.}
\label{Fig6} 
\end{figure}

The in-plane magnetic field of a two-carrier system may cause an electron-hole imbalance provided the sample width $d$ is smaller than the electron-hole recombination length $\ell_0$. Depending on impurity scattering and the details of electron-phonon coupling, the latter may vary in HgTe quantum wells from ten nanometers to a few microns. Thus, we expect that the condition $\ell_0 > d$ is indeed fulfilled in our experiment. In this case, one cannot regard the sample as truly three-dimensional since both charge and quasi-particle densities strongly vary in the $z$ direction (perpendicular to the plane) when a magnetic field is applied, even if the local density of states is constant everywhere in the sample. 
Assuming an idealized electron-hole symmetric system with the same mobilities $\mu_q$, the resistance at charge neutrality can be described by the following formula
\begin{equation}
\label{theory}
R_{xx,max}=R_0  \frac{1+(\mu_q B)^2}{1+(\mu_q B)^2 \frac{\tanh (d/\ell_R(B))}{d/\ell_R(B)}},
\end{equation}
where $B$ is the in-plane magnetic field (applied perpendicular to the current), $\ell_R(B)=\ell_0 \lt[1+(\mu_q B)^2\rt]^{-1/2}$ is the magnetic field-dependent recombination length and $\ell_0$ and $R_0$ are the values of the electron-hole recombination length and sample resistance at zero field respectively. 
More details about the theoretical derivation and description of the mechanism that gives rise to this magnetoresistance can be found in the Appendix.
Eq.~(\ref{theory}) already suggests a roughly linear upturn in the maximal resistance in strong fields $\mu_qB\gg \ell_0/d$ as shown in Fig.~\ref{Fig6}: an effect which is caused entirely by a classical drift of electrons and holes in a magnetic field. This effect takes place assuming no influence of the magnetic field on the impurity scattering rates, the density of states and the spectrum in general. Much sharper upturns in resistance, demonstrated in Fig.~\ref{Fig5}, may be caused by a combination of this classical effect with the direct suppression of quasi-particle mobility by the magnetic field. 

The theory presented in the Appendix also predicts that the width of the resistance peak in the gate voltage space grows with the ratio of the Thomas-Fermi screening length to the sample thickness $d$.
The large shift in the energy bands that causes the transition of 2D TI states to the semi-metallic phase in the first place leads to a strong overlap of the conduction and valence bands that increases with increasing magnetic field. This, in turn, leads to a two-liquid bulk transport that is strongly affected by the magnetic field as explained in the Appendix. Thus, we conclude that the observed upturn of resistance may originate, at least in part, from the semi-metalicity that is created once the sample thickness becomes smaller than the electron-hole recombination length.

According to the theory, the magnetoresistance in the semi-metallic phase is maximal at the charge compensation point, which is indicated here by the notation $R_\textrm{xx,max}$.
This is, as already shown, consistent with the experimental findings presented in Fig.~\ref{Fig4} where the strongest magnetoresistance response is observed close to charge neutrality.

We expect the classical mechanism of two-liquid magnetoresistance to be mainly responsible for the high-field behaviour at fields of 30\,T and above. It is, however, possible that additional effects, including a spectrum reconstruction and a magnetic field dependence of the scattering rate also contribute.  We emphasize, that the observed magnetoresistance is of classical nature and that we observe the same quantitative behaviour for samples with different quantum well thicknesses. 

In summary, we have shown that the previously observed phase transition from a 2D TI to a semi-metallic state can experimentally be observed in quantum wells with thicknesses between 7 and 11\,nm. High magnetic fields up to 33 T enable us to study the evolution of the magnetoresistance in this gapless state, and we observe for the first time the onset of an exponential divergence of the longitudinal resistance at fields of order 30 T in these systems.
This divergence is likely related to a magnetic-field-driven change in the quasi-particle density caused by the overlapping of conduction and valence bands. This feature occurs in all our samples independent of the well thickness. 
   
\begin{acknowledgments}
This work has been performed at the HFML-RU/FOM member of the European Magnetic Field Laboratory (EMFL) and is part of the research programme of the Foundation for Fundamental Research on Matter (FOM), which is part of the Netherlands Organization for Scientific Research (NWO). M.T. acknowledges the support from the Russian Science Foundation under the Project 17-12-01359. 
\end{acknowledgments}

\bibliography{Lit}

\appendix*
\section{Appendix: Magnetoresistance of the two-carrier system}
\begin{figure}
\includegraphics[width=\linewidth]{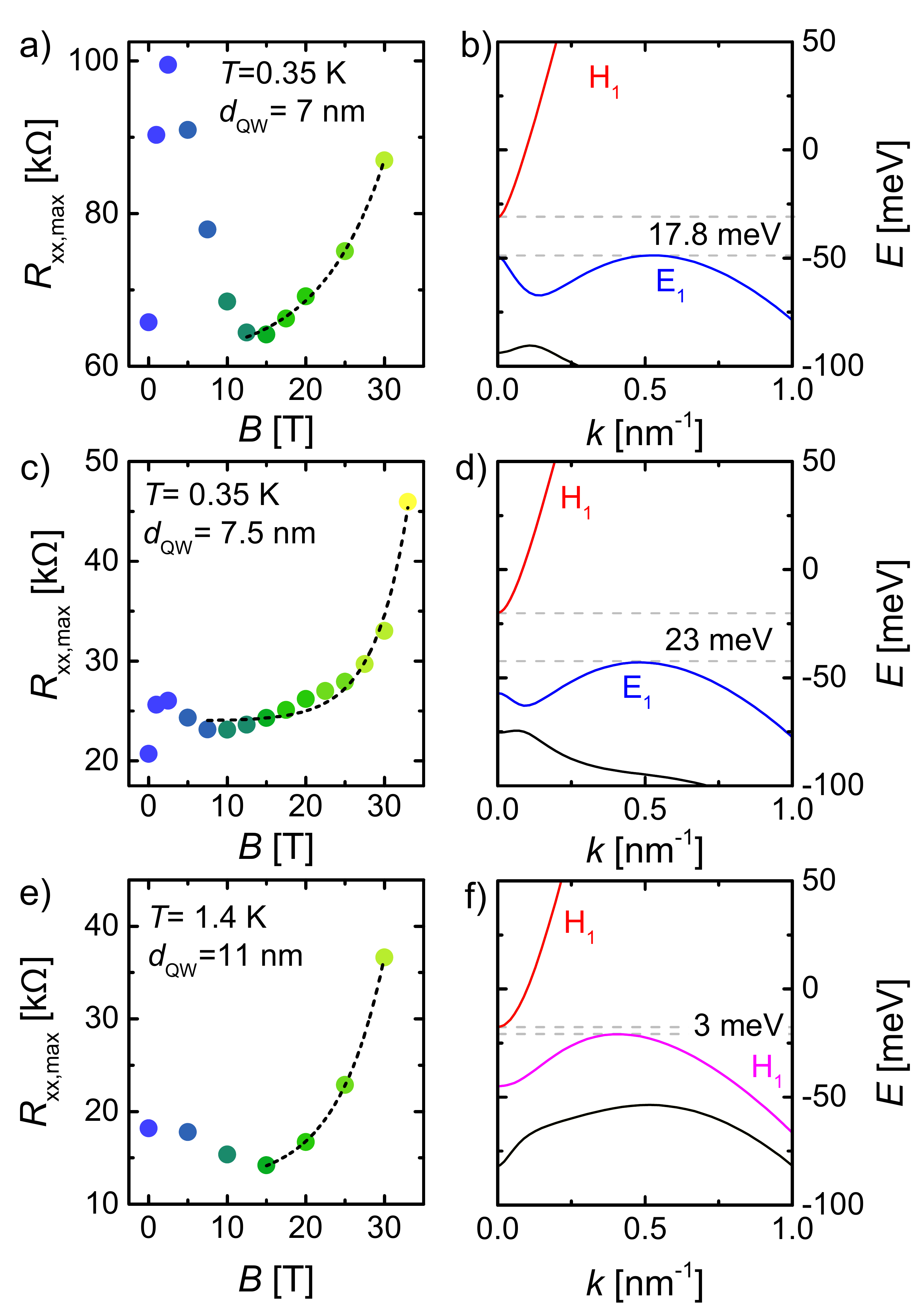}%
\caption{Peak value of the longitudinal resistance $R_{xx,max}$ for HgTe quantum wells with thicknesses of  a)$d_\textrm{QW}= 7$ nm, c) $d_\textrm{QW}= 7.5$ nm and e) $d_\textrm{QW}= 11$ nm. Black dashed lines are a guide for the eye that show the divergence of resistance at fields above 15 T. Panels b), d) and f) show the corresponding band structure calculations where conduction and valence bands are labelled by H$_1$ and E$_1$ (H$_2$), respectively. Bulk band gaps are indicated by grey dashed lines.}
\label{Fig5} 
\end{figure}

In Fig.~\ref{Fig5} a), c) and e) we show the data of $R_\textrm{xx,max}$ for the 7, 7.5 and 11\,nm samples, respectively.  On the left of the experimental plots in Fig.~\ref{Fig5} b),d) and f) we additionally provide $\mathbf{k\cdot p}$ calculations of the band structure in zero magnetic field. The samples show the same qualitative behaviour as the 9\,nm sample discussed in the main part of this publication.  In table\ref{Table1} we provide an overview of the samples that were used in this study.

\begin{table}
\centering
\begin{tabular}{|c|c|c|c|}
\hline 
Sample & $d_QW$ [nm] & Dimensions [$\mu m^2$]& $\Delta E$ [meV]  \\ 
\hline 
\# 1 & 7 & $1200\times 400$& 17.8\\ 
\hline
\# 2 & 7.5 & $600\times 200$& 23\\ 
\hline
\# 3 & 9 & $600\times 200$& 13\\ 
\hline
\# 4 & 11 & $600\times 200$& 3\\ 
\hline 
\end{tabular} 
\caption{\label{Table1} Summary of the samples used in this study. In the table the respective quantum well width $d_QW$, sample dimensions as well as the energy gaps $\Delta E$ at zero magnetic field as calculated from self-consistent  $k\cdot p$ calculations are provided}
\end{table}

In the following, we adopt the analysis of Ref.~\cite{Alekseev2017} to investigate the magnetoresistance of a thin three-dimensional sample in an in-plane magnetic field. 

As a simple example, we consider a semi-metal with symmetric electron and hole bands. We set zero energy at the band crossing such that the electron and hole dispersions are defined as
\be
\ep^\pm_{\bb{p}}=\pm\frac{p^2}{2m},
\e
where $m$ is an effective mass. Consequently, we define equilibrium electron and hole quasi-particle concentrations as
\be
\n
n^{(0)}_e=\int_0^\infty d\ep\,\nu(\ep) f(\ep),\quad n^{(0)}_h=\int_0^\infty d\ep\,\nu(\ep) (1-f(-\ep)),
\e
where $f(\ep)=\lt(1+\exp\lt[(\ep-\mu)/T\rt]\rt)^{-1}$ is the Fermi distribution function with the chemical potential $\mu$ and $\nu(\ep)$ is the 3D density of states
\be
\nu(\ep)=\frac{m}{\pi^2\hbar^3}\sqrt{2m\ep}.
\e
Thus, we restrict ourselves below to the case of perfect electron-hole symmetry. The generalization to more general forms of spectra is, however, straightforward. 

By integrating the Boltzmann kinetic equation, provided e.\,g. in Ref.~\cite{Alekseev2017}, we justify the so-called drift-diffusion equations for the symmetric two-liquid system of the form
\beml
\begin{align}
&-\frac{v^2}{3}\bb{\nabla}n_e+\frac{v^2}{3}\frac{\pa n_e}{\pa \mu}e\bb{E} - \bb{j}_e\times \bb{\omega}_c =\frac{\bb{j}_e}{\tau},\\
&-\frac{v^2}{3}\bb{\nabla}n_h+\frac{v^2}{3}\frac{\pa n_h}{\pa \mu} e\bb{E}+ \bb{j}_h\times\bb{\omega}_c  =\frac{\bb{j}_h}{\tau},
\end{align}
\eml
where $\bb{\omega}_c=e\bb{B}/mc$ is the vector cyclotron frequency, $\bb{B}$ is the magnetic field, $\bb{E}$ is the electric field, $v$ is the Fermi velocity, $\bb{j}_{e,h}$ are the quasi-particle electron and hole current densities, and $\tau$ is a transport scattering time. Multiplying the equation with $\tau$ and introducing the diffusion coefficient $D=v^2\tau/3$ we obtain
\beml
\label{dd}
\begin{align}
&\bb{j}_e=D\frac{\pa n^{(0)}_e}{\pa \mu} e\bb{E} - \bb{j}_e\times \bb{\omega}_c \tau -D\bb{\nabla}n_e,\\
&\bb{j}_h=D\frac{\pa n^{(0)}_h}{\pa \mu} e\bb{E} + \bb{j}_h\times \bb{\omega}_c \tau -D\bb{\nabla}n_h,
\end{align}
\eml
where we take advantage of equilibrium densities $n^{(0)}_{e,h}$ since the linear response with respect to electric field is assumed. 

The electric current density in the two-liquid model is defined as $\bb{J}=e(\bb{j}_e-\bb{j}_h)$, where $e$ is the electron charge (which is here negative). Similarly, we may define the quasi-particle current density as $\bb{P}=\bb{j}_e+\bb{j}_h$. In contrast to $\bb{J}$, the latter is not conserved and may flow in a different direction. Indeed, the quasi-particle currents are intrinsically decaying since electrons and holes may annihilate each other or be created due to inelastic processes, e.\,g. due to the interactions with phonons. 

We further introduce the charge density and the quasi-particle density as $n=n_e-n_h$ and $\rho=n_e+n_h$, respectively and denote the electron-hole recombination time (e.\,g. due to phonons) as $\tau_{R}$. The drift diffusion equations of Eqs.~(\ref{dd}) have to be supplemented with  equations for the currents, that may also be derived by integrating of the corresponding Boltzmann kinetic equation,
\be
\label{currents}
\bb{\nabla}\cdot\bb{j}_e=-\frac{\rho-\rho_0}{2\tau_{R}},\qquad \bb{\nabla}\cdot\bb{j}_h=-\frac{\rho-\rho_0}{2\tau_{R}},
\e 
where $\rho_0=n_e^{(0)}+n^{(0)}_h$ is the quasiparticle concentration in equilibrium (i.\,e. in the absence of currents). This system of equations (\ref{currents}) must also be supplemented with the Poisson equation for a 3D sample,
\be
\bb{\nabla}\cdot \bb{E}=4\pi e\,n.
\e 
The equations above must now be solved in a finite 3D geometry with appropriate boundary conditions that correspond to the absence of currents across the closed boundaries with a vanishing component of the electric field parallel to the sample interface. We will not undertake a full 3D solution assuming that the in-plane sample dimensions are essentially infinite when compared to all electronic length scales.  Therefore, it is sufficient to seek a solution that is translationally invariant in the $xy$ plane.

The equations above can be rewritten in terms of quasi-particle and electric currents as
\beml
\label{eqs1}
\begin{align}
&\bb{J}=\sigma_0 \bb{E}+e\beta\, \bb{b}\times \bb{P}-eD\bb{\nabla}n,\\
&e\bb{P}=\sigma_1 \bb{E} +\beta\, \bb{b}\times \bb{J}-eD\bb{\nabla}\rho,
\end{align}
\eml
where $\bb{b}$ is the unit vector in the direction of the in-plane magnetic field, $\beta=\omega_c\tau$. We also introduce
\be
\sigma_0=e^2D \frac{\pa n_0}{\pa\mu},\qquad \sigma_1=e^2D \frac{\pa \rho_0}{\pa\mu},
\e
where $n_0=n_e^{(0)}-n^{(0)}_h$. Note that $\sigma_0$ is positive definite (since $n_0$ is an odd monotonous function of $\mu$). This is nothing but the sample conductivity in the absence of the magnetic field. The quantity $\sigma_1$ changes sign at charge neutrality (since $\rho_0$ is an even function of $\mu$) while it approaches $\pm \sigma_0$ away from charge neutrality. Thus, we always have $|\sigma_1| \leq \sigma_0$. In the semi-metal regime we can assume $\sigma_1\ll \sigma_0$, while the resistance reaches its maximal value $R_\textrm{xx,max}$ at charge neutrality, which corresponds here to $\sigma_1=0$.

The remaining equations are conveniently written as 
\be
\label{eqs2}
\bb{\nabla}\bb{J}=0,\quad \bb{\nabla}\bb{P}=-(\rho-\rho_0)/\tau_{R},\quad \bb{\nabla}\bb{E}=4\pi e\,n,
\e
where the first one expresses the current conservation, the second the quasi-particle current decay, and the third is the Maxwell equation that defines the electrostatics of the sample.

We apply the equations to a semi-metal rectangular sample of dimensions $(L,W,d)$ where $d\ll L, W$. The current is injected along $x$ direction, $0<x<L$. We assume that the system is translationally invariant in the $x$ and the $y$ directions and take only the possible $z$ dependence of all quantities into account. In this case, the current conservation reads $\pa J_z/\pa z=0$ hence $J_z=0$ due to boundary conditions on current in $z$ direction. We also apply the magnetic field in-plane such that $\bb{b}=(b_x,b_y,0)$.
Thus, the system of equations (\ref{eqs1}) is reduced to
\beml
\label{eqs3}
\begin{align}
&J_x=\sigma_0 E_x+e\beta b_y P_z,\\
&J_y=\sigma_0 E_y-e\beta b_x P_z,\\
&0=\sigma_0 E_z+e\beta (b_x P_y-b_y P_x)-e D \pa_z n,\\
&e P_x = \sigma_1 E_x,\\
&e P_y = \sigma_1 E_y,\\
&e P_z = \sigma_1 E_z+\beta(b_xJ_y-b_yJ_x)-eD\pa_z\rho.
\end{align}
\eml
These equations are combined with Eqs.~(\ref{eqs2}). Since we have already assumed  translational invariance in the $x,y$ plane, we must solve the system for constant ($z$-independent) $E_x$ and $E_y$ to obtain the profiles $J_x(z)$ and $J_y(z)$. The system is, then, viewed as a set of parallel two-dimensional planes with different two-dimensional conductivity tensors $\hat{\sigma}(z)$. These conductivities naturally sum up hence we should simply integrate the tensor over the $z$ coordinate and invert it to obtain the resistivity. 

The subsequent solution is straightforward. From Eqs.~(\ref{eqs3}) we obtain the pair of equations
\beml
\label{eqs4}
\begin{align}
\label{Ez}
&\sigma_0 E_z-e D\pa_z n +\beta \sigma_1 (b_x E_y-b_y E_x)=0,\\
&eP_z=\frac{\sigma_1}{1+\beta^2}E_z-\frac{eD}{1+\beta^2}\pa_z\rho+\frac{\beta\sigma_0}{1+\beta^2} (b_x E_y-b_y E_x)
\label{Pz}
\end{align}
\eml
Differentiating Eq.~(\ref{Ez}) over $z$ and using Eq.~(\ref{eqs2}) we obtain the standard equation for the screening
\be
\frac{\pa^2 n}{\pa z^2}=4\lambda^2 n, \qquad \lambda=\sqrt{\pi \sigma_0/D},
\e
where $\lambda$ is the inverse Thomas-Fermi length. The solution to this equation is substituted back into Eq.~(\ref{Ez}) to obtain the function $E_z(z)$. The freedom of the solution is fixed by the condition $E_z(\pm d/2)=0$.

Differentiating the second equation (\ref{Pz}) and using Eq.~(\ref{eqs2}) we obtain the equation on the quasi-particle density
\be
\label{rho}
\frac{\pa^2 \rho}{\pa z^2}=4\kappa^2 (\rho-\rho_0)+\frac{4\pi\sigma_1}{eD} \frac{\pa E_z}{\pa z}, 
\e
where $\kappa=\sqrt{(1+\beta^2)/4D\tau_R}$ is the inverse recombination length (this length is inversely proportional to the magnetic field for $\beta=\omega_c\tau\gg 1$). The equation of Eq.~(\ref{rho}) is readily solved and the solution can be substituted into Eq.~(\ref{Pz}) to obtain $P_z(z)$. The freedom of the solution is fixed by the condition $P_z(\pm d/2)=0$.

At the next step, we integrate the first two equalities in Eq.~(\ref{eqs3}) over $z$ from $-d/2$ to $d/2$ to obtain corresponding currents (per system width $W$). These are proportional to the electric fields $E_x=V_x/L$ and $E_y=V_y/L$, where $V_x$ and $V_y$ are the corresponding voltages. Inverting the corresponding tensor, we relate the voltage $V_x$ to the total current $I_x$ to define the longitudinal resistivity.

The result of this straightforward calculation is rather cumbersome but can still be written for arbitrary in-plane direction of magnetic field as
\begin{align}
R_{xx}&=R_0(1+\beta^2)\Bigg[1+\beta^2\lt(1+b_y^2\lt[U(\kappa d,\lambda d)-F(\kappa d)\rt]\rt)\n\\
&-\frac{b_x^2b_y^2\beta^4\lt[U(\kappa d,\lambda d)-F(\kappa d)\rt]^2}{1+\beta^2\lt(1+b_y^2\lt[U(\kappa d,\lambda d)-F(\kappa d)\rt]\rt)}\Bigg]^{-1} 
\label{final}
\end{align}
where $R_0=L/Wd\sigma_0$. We have also defined the functions
\be
\n
U(x,y)=\frac{\sigma_1^2}{\sigma_0^2}\frac{x^2F(y)-y^2 F(x)}{x^2-y^2},\quad
F(x)=1-\frac{\tanh(x)}{x}.
\e
Note that this simple solution predicts zero magnetoresistance for $B$ applied in the direction of charge transport, i.\,e. for $b_y=0$. 

It is easy to see that there are three different length scales in the solution of the Eq.~(\ref{final}), namely the system thickness $d$, the recombination length $\ell_R=\kappa^{-1}=\ell_0/\sqrt{1+\beta^2}$, where $\ell_0=2\sqrt{D\tau_R}$, and the Thomas-Fermi screening length $\ell_{TF}=1/\lambda$. The final result is particularly simple for $\ell_{TF}\gg d$ and for the magnetic field directed along the$y$ axis (perpendicular to the current). In this case, we find 
\be
\label{simple}
R_{xx}= R_0 \frac{1+\beta^2}{1+\beta^2 \frac{\tanh d/\ell_R}{(d/\ell_R)}}.
\e
The same expression is recovered at charge neutrality $\sigma_1=0$ where the resistance reaches its maximal value. The solution to Eq.~(\ref{simple}) corresponds to Eq.~(\ref{theory}) of the main text. We see that the result acquires a strong magnetic field dependence in the regime $\ell_R \ll d$, i.\,e. for large magnetic fields. In the opposite limit of a very thin sample, $d\ll \ell_R$ we recover naturally the field-independent resistance $R_{xx}= R_0$. 

It is also not difficult to write down the result for arbitrary $\ell_{TF}=1/\lambda$ (with the magnetic field still directed along the $y$ axis) as
\be
\n
R_{xx}= \frac{R_0(1+\beta^2)}{1+\beta^2 \frac{\tanh (\kappa d)}{\kappa d} + \beta^2\frac{\sigma_1^2}{\sigma_0^2}\lt(1+\frac{\lambda^3\tanh(\kappa d)-\kappa^3\tanh(\lambda d)}{(\kappa^2-\lambda^2)\kappa \lambda d}\rt)}.
\e
Another interesting limit is the limit of large sample size such that both $d\lambda$ and $d\kappa$ are large parameters, i.\,e. in the regime $d\gg \ell_{TF},\ell_R$.
In this limit one obtains
\be
R_{xx}= \frac{R_0}{1-\frac{\beta^2}{1+\beta^2} b_y^2\eta^2 \lt(1+\frac{\eta^2b_x^2\beta^2}{1+\beta^2(1-b_y^2\eta^2)}\rt)}, \qquad 
\e
where
\be
 \eta^2=1-\frac{\sigma_1^2}{\sigma_0^2}.
\e
 
\end{document}